# Rogue waves in a two-component Manakov system with variable coefficients and an external potential

Wei-Ping Zhong[1,2] [*], Milivoj Belić[2], and Boris A. Malomed[3]

[1] *Department of Electronic and Information Engineering, Shunde Polytechnic, Guangdong Province, Shunde 528300, China*

[2] *Texas A&M University at Qatar, P.O. Box 23874 Doha, Qatar*

[3] *Department of Physical Electronics, School of Electrical Engineering, Faculty of Engineering, Tel Aviv University, Tel Aviv 69978, Israel*

*Corresponding author: zhongwp6@126.com

**Abstract:** We construct rogue waves (RWs) in a coupled two-mode system with the self-focusing nonlinearity of the Manakov type (equal SPM and XPM coefficients), spatially modulated coefficients, and a specially designed external potential. The system may be realized in nonlinear optics and Bose-Einstein condensates. By means of a similarity transformation, we establish a connection between solutions of the coupled Manakov system with spatially-variable coefficients and the basic Manakov model with constant coefficients. Exact solutions in the form of two-component Peregrine and dromion waves are obtained. The RW dynamics is analyzed for different choices of parameters in the underlying parameter space. Different classes of RW solutions are categorized by means of a naturally introduced control parameter which takes integer values.

**PACS number(s):** 05.45.Yv, 42.65.Tg, 42.65.Sf, 02.30.Ik

## I. Introduction

Rogue waves (RWs), or freak waves, are nonlinear excitations in the ocean that exhibit abnormal height for a short period of time [1-4]. A RW typically has amplitude about three times larger than surrounding waves, "appear from nowhere and disappear without a trace" [5]. The RW concept of giant ocean waves has been extended to other physical settings, such as plasmas [6], nonlinear light propagation in doped fibers [7,8] and other optical media [9,10], acoustics [11], microwave transport [12], and Bose-Einstein condensates (BECs) [13].

The dynamics of RWs is well modeled by the nonlinear Schrödinger equation (NLSE) (see Refs. [7-15] and references therein) and the derivative NLSE [16], which give rise to solutions featuring localized spatiotemporal regions with a high amplitude, relative to the background. The formation of such high-amplitude wave packets is fueled by the modulation instability of the background [17, 18]. However, such special solutions do not necessarily represent RWs *per see*. The analysis requires to study the propagation of such waves, allowing them to change and collide, and treat the results statistically, by determining from long-tail intensity distribution histograms what percentage of the observed waves may be categorized as real RWs [7-10]. In this paper, we follow the broad definition of RWs, as special analytical solutions of NLSE supported by a finite background – such as Peregrine, Akhmediev, or Ma breathers, that may represent real rogue waves in experimental settings.

Many complex systems may involve more than one component, which can be related to the dispersive and gain/loss properties of the system [19]. In this connection, the study of RWs was extended to two-component systems [20-24], where new excitations with noteworthy features have been predicted numerically and analytically [25-27]. Additional novel RW patterns produced by multi-component models have been reported in Refs. [28-36]. In particular, the following types of RG solutions have been obtained in the system of two coupled NLSEs with constant coefficients: Breathers and rational-breather solutions of multi-component NLSEs were presented in Ref. [25] in a determinant form, as degenerate limits of available algebraic-geometric solutions. Explicit RW solutions of two coupled NLSEs have been constructed in Refs. [20,28,37] by dint of the modified Darboux transformation. Basic properties of multi-RW solutions and their collision structures have also been studied in Ref. [37].

In all the above-mentioned works, explicit forms of RW solutions of the coupled NLSEs with constant coefficients were explored. However, inhomogeneous nonlinear media are modeled by NLSEs in which coefficients are not constant. In addition, the traditional approach to RW propagation usually admits the dependence of diffraction and nonlinearity coefficients on the propagation distance only, which corresponds to the well-known concepts of dispersion and nonlinearity management [6,7,38]. The drawback of this approach is that it does not take into account transverse inhomogeneity and external potentials. Here, we address the transverse inhomogeneity and introduce a more general concept of a discrete-parameter control, which may include the diffraction and nonlinearity management.

In this paper, we present RW solutions of two NLSEs with the Manakov nonlinearity, *i.e.*, equal SPM and XPM (self- and cross-phase-modulation) coefficients [39], with spatially modulated coefficients and a specially devised external potential, with the help of a similarity transformation. In that context, there appears a discrete control parameter, which categorizes different types of vector RWs, such as the two-component Peregrine and dromion waves. Based on these solutions, we find that the coupled RWs can be easily managed by means of a properly chosen set of free coefficients.

The rest of the paper is structured as follows. In Sec. II we introduce the coupled NLSEs and convert them into the standard coupled NLSE system, using the similarity transformation, which makes it possible to find analytical RW solutions. In Sec. III we present different classes of solutions produced by the coupled NLSEs. Several typical examples of two-component RWs are analyzed in that section. In Sec. IV, we summarize the results.

## II. The model and vector RW solutions

We consider the scaled system of coupled NLSEs, following Ref. [15]:

$$i\frac{\partial u_1}{\partial z} + \beta(x)\frac{\partial^2 u_1}{\partial x^2} + 2\chi(x)\left(|u_1|^2 + |u_2|^2\right)u_1 + U(x)u_1 = 0, \quad (1a)$$

$$i\frac{\partial u_2}{\partial z} + \beta(x)\frac{\partial^2 u_2}{\partial x^2} + 2\chi(x)\left(|u_1|^2 + |u_2|^2\right)u_2 + U(x)u_2 = 0, \quad (1b)$$

where $u_1(z,x)$, $u_2(z,x)$ are complex wave amplitudes, $z$ and $x$ represent the propagation distance and the transverse coordinate, function $\beta(x)$ represents the effective diffraction coefficient, and $\chi(x)$ is the nonlinearity coefficient. This setting naturally arises in models of planar optical waveguides subject to appropriate transverse modulation that can be implemented by means of $x$-dependent inner structure of the waveguide (including the ones built as arrays or lattices) and dopant density [40]. Equations (1), with $z$ replaced by time, may also be realized as coupled Gross-Pitaevskii equations for a nearly one-dimensional two-component BEC, in which case the $x$-dependence of the effective mass, $m_{eff}=1/2\beta(x)$, may be induced by means of a nonuniform optical lattice, combined with the cigar-shaped trapping configuration [41], while the $x$-dependent nonlinearity coefficient $\chi(x)$, can be induced by the appropriately shaped dc magnetic field [42]. Note that we consider the case when the coefficients depend on the transverse coordinate, rather than the propagation distance, which is more usual in the models of "management" [38].

Further, we assume the special external potential in a parabolic form, modulated by the same diffractive coefficient introduced in Eqs. (1):

$$U(x) = \beta(x)\left(ax^2 + b\right), \quad (2)$$

where $a$ and $b$ are two real constants, to be determined below. Following Ref. [15], we search for a similarity transformation that would reduce (1) to the standard Manakov system with constant coefficients:

$$i\frac{\partial V_1}{\partial z} + \frac{\partial^2 V_1}{\partial Y^2} + 2\left(|V_1|^2 + |V_2|^2\right)V_1 = 0, \quad (3a)$$

$$i\frac{\partial V_2}{\partial z} + \frac{\partial^2 V_2}{\partial Y^2} + 2\left(|V_1|^2 + |V_2|^2\right)V_2 = 0, \quad (3b)$$

where the complex fields $V_j = V_j(z,Y)$ ($j=1,2$) are functions of the propagation distance and the new similarity variable

$Y = Y(x)$. This approach, based on the generation of seemingly complex models by transformation of known integrable equations may seem somewhat artificial, but, nevertheless, it is known to generate quite nontrivial results [43].

Solutions of the integrable Manakov system (3) have been obtained in various forms, including those produced by the Darboux transformation [20-37]. In particular, the latter method yields RW solutions [20, 28]:

$$\begin{bmatrix} V_1(z,Y) \\ V_2(z,Y) \end{bmatrix} = \begin{pmatrix} \dfrac{k_1 L + k_2 M}{B} \\ \dfrac{k_2 L - k_1 M}{B} \end{pmatrix} e^{2i\Omega z}, \quad (4a)$$

$$L = \frac{3}{2} - 8\Omega^2 z^2 - 2k^2 Y^2 + 8i\Omega z + |f|^2 e^{2kY}, \quad (4b)$$

$$M = 4f\left(kY - 2i\Omega z - \frac{1}{2}\right) e^{kY + i\Omega z}, \quad (4c)$$

$$B = \frac{1}{2} + 8\Omega^2 z^2 + 2k^2 Y^2 + |f|^2 e^{2kY}. \quad (4d)$$

Here $k_1$ and $k_2$ are arbitrary real parameters, the spatial frequency is $\Omega = k^2 = k_1^2 + k_2^2$, and $f$ is an arbitrary complex constant [20]. Following the procedure developed in Refs. [15,16,44], we seek for the similarity transformation of Eq. (1) into (3), with the form of the solution

$$u_j(z,x) = A(x) V_j(z,Y), \quad (5)$$

where $A(x) > 0$ is a real amplitude, and the similarity variable $Y = Y(x)$ is to be determined. The substitution of Eq. (5) into (1) indeed leads to Eq. (3), provided that a system of differential equations for $Y(x)$ and $A(x)$ is satisfied,

$$2A_x Y_x + A Y_{xx} = 0, \quad (6a)$$

$$\beta A_{xx} + UA = 0, \quad (6b)$$

in which the subscript stands for derivative, while the nonlinearity and diffraction coefficients in Eq. (1) are given by

$$\chi(x) = [A(x)]^{-2}, \quad \beta(x) = \left(\frac{dY}{dx}\right)^{-2}. \quad (7)$$

Equation (6a) can be solved immediately,

$$Y(x) = \int [A(x)]^{-2} dx. \quad (8)$$

In Eq. (2), real constants $a$ and $b$ may take different values. Here, we choose

$$a = -1/4, \quad b = m + 1/2, \quad (9)$$

to ensure that the external parabolic potential in the combination with the diffraction coefficient leads to a solvable differential equation for amplitude $A(x)$ and eventually, provides an exact solution, which is a finding of obvious interest. Thus, when $m$ is a non-negative integer, from Eqs. (2) and (6b) one easily finds that amplitude $A$ obeys the standard Weber differential equation, $A_{xx} + \left(m + \dfrac{1}{2} - \dfrac{1}{4}x^2\right) A = 0$, whose solutions are the parabolic-cylinder functions, $D_m(x)$; hence the solution for $A$ can be written as $A(x) = \lambda D_m(x)$ [45], where $\lambda = 1/\left(\sqrt{2\pi} m!\right)$ is the respective normalization constant. The integer $m$ is referred to as the discrete control parameter. It provides a possibility to conveniently classify the solutions of the NLSE system in terms of solutions to the linear second-order ordinary differential equation, which is not obvious from a direct consideration of Eq. (6b).

Finally, from Eqs. (4) and (5), we obtain the following analytical solution of Eq. (1):

$$\begin{bmatrix} u_1(z,x) \\ u_2(z,x) \end{bmatrix} = \frac{1}{\sqrt{2\pi m!}} D_m(x) \begin{bmatrix} V_1(z,Y) \\ V_2(z,Y) \end{bmatrix}, \quad (10)$$

where $V_j(z,Y)$ and $Y(x)$ are determined by Eqs. (4) and (8), respectively. Because the RW is given by Eq. (4) as a localized wave packet, it follows from Eq. (10) that $|u_j(z,x)| \to$ constant at $|x| \to \infty$, hence this solution is localized on top of a flat background. It should be stressed that, although Eqs. (1a) and (1b) are obviously symmetric with respect to the two components, the above RWs may have different components.

## III. Analysis of the solutions

The RW solutions to coupled NLSEs (3) with constant coefficients, for different parameters $k_1$, $k_2$ and $f$, were constructed in Ref. [20]. We utilize those wave forms to construct solutions given by Eq. (10) with different values of parameter $m$. Below, $I_{1,2} = |u_{1,2}|^2$ denote intensities of the two components of RWs.

The shape of potential $U(x)$, as defined by Eqs. (2) and (9), is displayed in Fig. 1 for $m = 0,1,2$. It is seen that it has $m+2$ zero points and $m+1$ local extrema. Such a potential can be realized experimentally and theoretically [46].

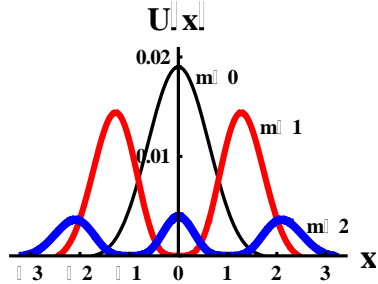

Fig. 1 (Color online) External potential $U(x)$ for different integers $m$, as per Eqs. (2) and (9).

The simplest solution is the RW which corresponds to a single-component wave form reported in Refs. [15,47]. It should be noted that, for $k_2 = 0$ and $f = 0$ in Eqs. (4) and (9), there exists only the $u_1$ component, while $u_2$ vanishes. In this case, the vector RW solution (10) is actually a scalar RW. Another special case is obtained when we set only $f = 0$, hence $M = 0$ in Eq. (4c). From Eqs. (4) and (10), one then finds that $u_1(z,x)$ is proportional to $u_2(z,x)$. In this way, a two-mode Peregrine soliton [48], with the two components proportional to each other, is obtained. In Fig. 2, we display the intensity distributions of such RWs.

For $m = 0$, the single Peregrine soliton was reported in Ref. [20]. For $m > 0$, multiple Peregrine solitons are found. The top row in Fig. 2 shows a two-peak RW for $m = 1$, with the amplitude of the $u_1$ component larger than that of $u_2$. For $m = 2$, a three-peak Peregrine RW appears in the middle row of Fig. 2, with the central peak lower than the other two. The bottom row of Fig. 2 exhibits a four-peak Peregrine RW with $m = 3$. Generally, the solution of order $m$ includes $m+1$ peaks, represented by the Peregrine RWs, along the $z = 0$ line. For odd $m > 0$, the intensity is zero at the origin, $(z,x) = (0,0)$, whereas for even $m$, the intensity forms a peak at this point. In the limit of large $m$, one obtains the two-component RW akin to the transversely modulated Akhmediev breather.

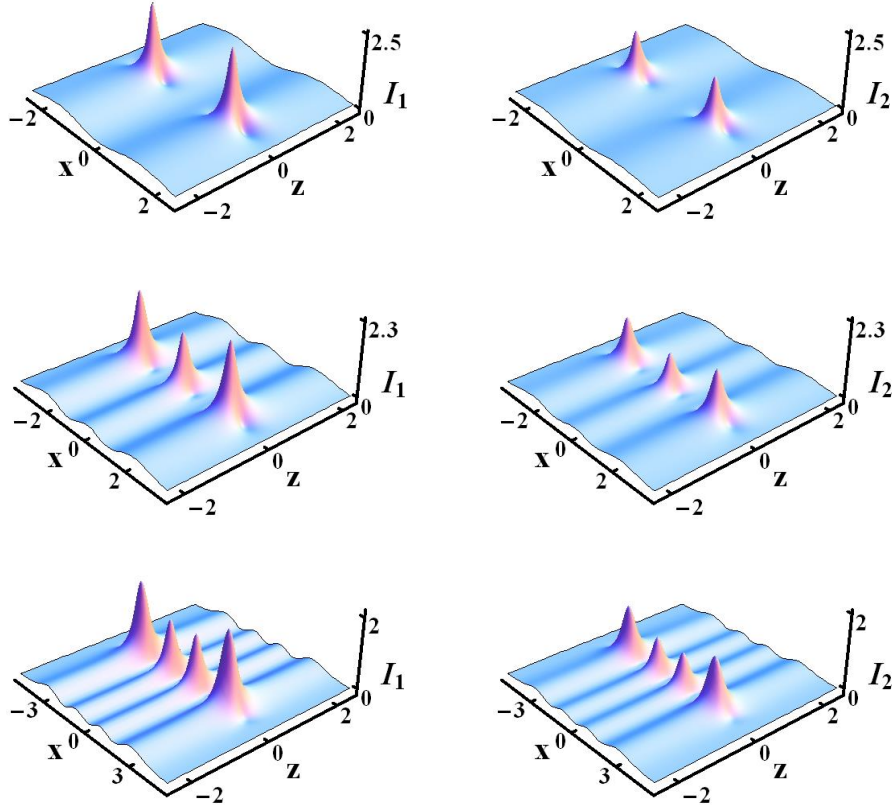

Fig. 2 (Color online) Two-mode Peregrine RWs with $k_1 = 1$, $k_2 = 0.8$, and $f = 0$, for $m = 1, 2, 3$ from top to bottom.

Novel RWs in the coupled NLSE system (1) are produced by setting $f \neq 0$ in the analytical solution (10), which corresponds to the so-called dromion solutions. These are two-dimensional localized patterns in the $(x, z)$ plane, produced by intersections of quasi-one-dimensional stripes (or "tracks", the Greek name for which, "dromos", gave the name to this class of solutions) [49]. A typical example is obtained for $k_2 = 0$, which enforces a possibility of having two unequal components, in the solution. Therefore, below we focus on effects of nonzero constant $f$ in Eq. (10).

In Fig. 3, we plot examples of the coupled RW solution (10) with a large real constant $f \gg 1$, generated by different $m$, which display a typical dromion structure [50]. With the increase of $m$, the number of dromions also increases. In general, for different $m$, we find that the solution is built of $m$ dromions.

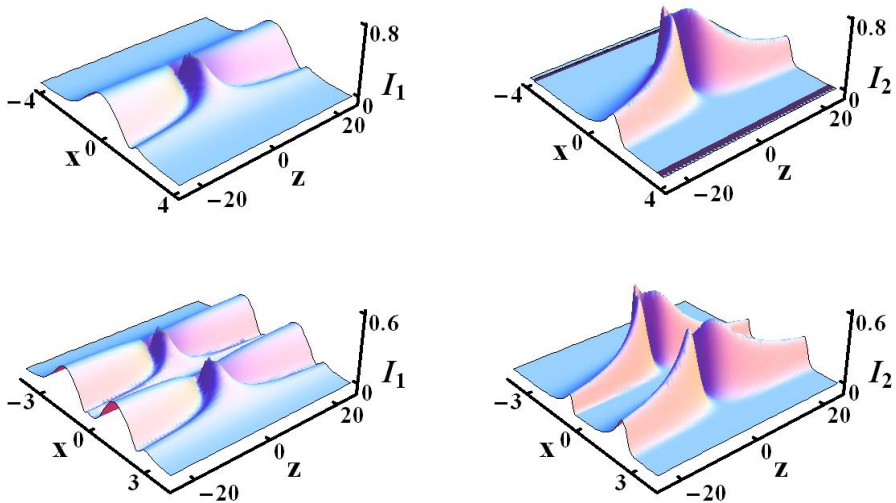

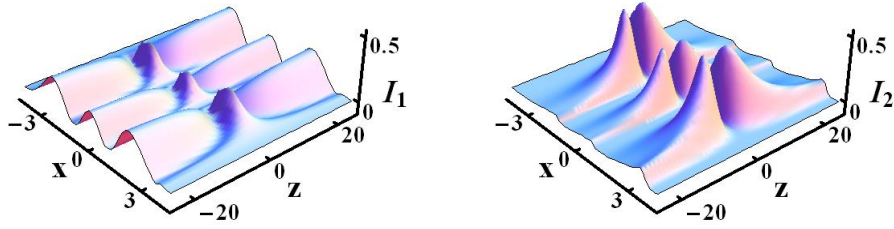

Fig. 3 (Color online) Profiles of the vector dromion RWs produced by solution (10) with $k_1 = 1$, $k_2 = 0$, and $f = 8$, for $m = 1,2,3$ from top to bottom.

Next, we discuss the vector RW solution (10) with the smaller value of the constant $f$, $0 < f < 1$. In Fig. 4, we plot a dromion with $k_1 = 1$, $k_2 = 0$, and $f = 0.8$, for different $m$. It is seen that for $m = 0$, the profile of the dromion wave $u_1$ appears as a bell-shaped peak on top of a long crest extended along the z-direction, whereas $u_2$ displays a peak with a dip. The peak is located at the origin $(z, x) = (0,0)$. As $m$ increases, these dromions become multi-peak structures, as seen in the middle and bottom rows of Fig. 4.

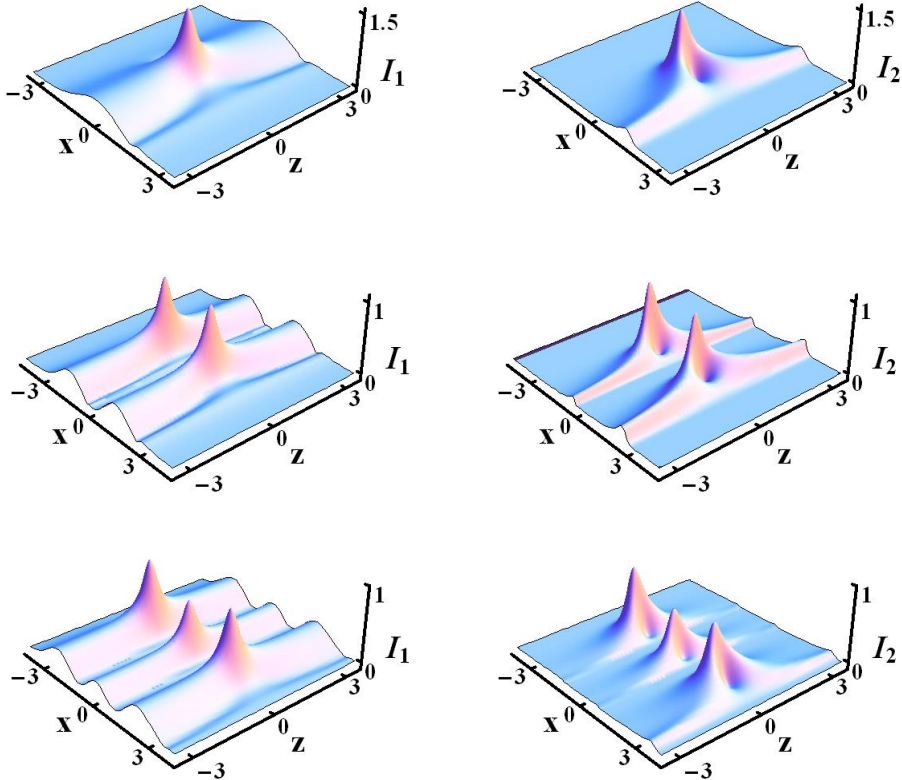

Fig. 4 (Color online) Profiles of the vector dromion RWs with $k_1 = 1$, $k_2 = 0$, and $f = 0.8$, for $m = 0,1,2$ from top to bottom.

Finally, when $k_1 \neq 0$, $k_2 \neq 0$, and $f \neq 0$, we obtain vector RWs that are made of Peregrine-like solitons sitting on top of long crests elongated along $x$-axis. Figure 5 illustrates the structure of these RWs for $k_1 = 1.2$, $k_2 = 1$, and $f = 0.04i$. The figure depicts the change in the beam intensity when the parameter $f$ changes from a real to an imaginary value. In the top row of Fig. 5, with $m = 0$, the peaks of the $u_1$ and $u_2$ components are located at $(z, x) = (0,0)$. By increasing $m$ to 1, a nonlinear wave packet with two peaks is generated, while the intensity vanishes at $(z, x) = (0,0)$, as seen in the bottom row of Fig. 5. Note also the development of additional structures in the transverse direction. This feature is not unusual for analytical

RW solutions [51]. In fact, when extended propagation along the *z*-direction is considered, all these waves become modulationally unstable (as any RW would do), with the development of complicated nonlinearly interfering wave fronts, that create conditions for the observation and statistical analysis of various other RWs. This research direction and the analysis is more in line with the standard consideration of RWs in nonlinear optics [52], which is beyond the scope of the present paper.

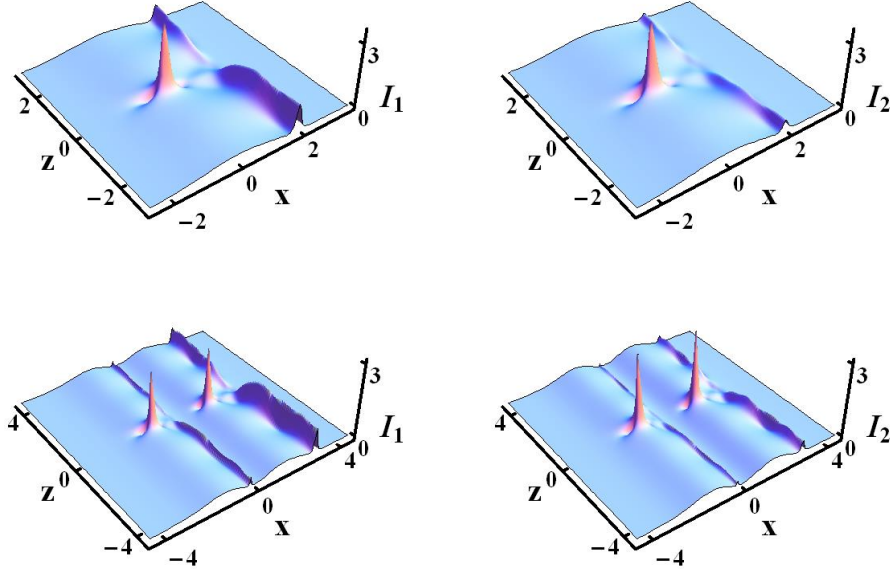

Fig. 5 (Color online) Evolution of the vector RW with $k_1 = 1.2$, $k_2 = 1$, $f = 0.04i$ for, $m = 0$ and $m = 1$, in the top and bottom rows.

## IV. Conclusion

We have considered the generalized two-mode coupled NLSE system in inhomogeneous media with the Manakov nonlinearity and the specially designed potential, by means of the similarity transformation, which reduces the system to the standard Manakov system. By choosing the special form of the effective diffraction coefficient, a simple procedure is established to obtain different classes of exact rogue wave modes. Using this procedure, we have found that the novel rogue waves are classified by the value of the integer control parameter $m$. These results may be helpful in finding new ways to manipulate RWs in inhomogeneous bimodal nonlinear media.

This work was supported, in a part, by the National Natural Science Foundation of China under grant No. 61275001 and by the Natural Science Foundation of Guangdong Province, China, under Grant No. 2014A030313799. The work at Texas A&M University at Qatar was supported by the NPRP 6-021-1-005 project with the Qatar National Research Fund (a member of Qatar Foundation).